\documentclass[pdftex]{article}
\usepackage{color}
\usepackage[dvipdfmx]{graphicx}       
\begin{document}

\begin{center}{\Large An implimentation of the Differential Filter\\ 
for Computing Gradient and Hessian \\[2mm]
of the Log-likelihood of Nonstationary Time Series Models}\\

\vspace{5mm}
{\large Genshiro Kitagawa}\\
Mathematics and Informatics Center, The University of Tokyo

\vspace{3mm}
{\today}

\end{center}

\noindent{\bf Abstract}

The state-space model and the Kalman filter provide us with unified and computationaly 
efficient procedure for computing the log-likelihood of the diverse type of time series models.
This paper presents an algorithm for computing the gradient and the Hessian matrix of the log-likelihood
by extending the Kalman filter without resorting to the numerical difference.
Different from the previous paper \cite{Kitagawa 2020b}, 
it is assumed that the observation noise variance $R=1$. 
It is known that for univariate time series, by maximizing the log-likelihood of this restricted model,
we can obtain the same estimates as the ones for the original state-space model.
By this modification, the algorithm for computing the gradient and the Hessian becomes
somewhat complicated. However, the dimension of the parameter vector is reduce by one
and thus has a significant merit in estimating the parameter of the state-space model
especially for relatively low dimentional parameter vector.
Three examples of nonstationary time seirres models, i.e., trend model, statndard 
seasonal adjustment model and the seasonal adjustment model with AR componet are presented to
exemplified the specification of structural matrices.
\\


\section{Introduction: The Maximum Likelihood Estimation of a State-Space Model}

We consider a linear Gaussian state-space model 
\begin{eqnarray}
 x_n &=& F_n(\theta) x_{n-1} + G_n(\theta) v_{n} \label{Eq-SSM-1}\\
 y_n &=& H_n(\theta) x_n + w_n,  \label{Eq-SSM-2}
\end{eqnarray}
where $y_n$ is a one-dimensional time series, $x_n$ is an $m$-dimensional state vector, $v_n$ is a $k$-dimesional Gaussian white noise, $v_n \sim N(0,Q_n(\theta ))$, and $w_n$ is a one-dimensional white noise, $w_n \sim N(0,1)$.
$F_n(\theta )$, $G_n(\theta )$ and $H_n(\theta )$ are $m\times m$ matrix, $m\times k$ matrix and $m$ vector. respectively.
$\theta$ is the $p$-dimensional parameter vector of the state-space model such as the variances of the noise
inputs and unknown coefficients in the matrices $F_n(\theta )$, $G_n(\theta )$, $H_n(\theta )$ and $Q_n(\theta )$.
For simplicity of the notation, hereafter, the parameter $\theta$ and the suffix $n$ will be omitted.
It is noted that for the state-space model of the univariate time series, the assumption that $R=1$ does not
lose any genrality, since it is known that even with this assumption we can obtaine the same estimates of the
parameters of the model by a proper transformation (Kitagawa (2020)).

Various models used in time series analysis can be treated uniformly within the state-space model framework. 
Further, many problems of time series analysis, such as prediction, signal extraction, decomposition, parameter estimation and interpolation, can be formulated as the estimation of the state of a state-space model.

Given the time series $Y_N\equiv \{y_1,\ldots ,y_N\} $ and the state-space model (\ref{Eq-SSM-1})  
and (\ref{Eq-SSM-2}), the one-step-ahead predictor $x_{n|n-1}$ and the filter $x_{n|n}$ and their variance covariance 
matrices $V_{n|n-1}$ and $V_{n|n}$ are obtained by the following Kalman filter (Anderson and Moore (2012) and Kitagawa (2020)):

One-step-ahead prediction
\begin{eqnarray}
 x_{n|n-1} &=& F x_{n-1|n-1} \nonumber \\
 V_{n|n-1} &=& F V_{n-1|n-1} F^T + G Q_{n} G^T. \label{Eq-3-2}
\end{eqnarray}
\indent
Filter 
\begin{eqnarray}
 K_n &=& V_{n|n-1}H^T (H V_{n|n-1} H^T + R)^{-1} \nonumber \\
 x_{n|n} &=& x_{n|n-1} + K_n (y_n -H x_{n|n-1}) \label{Eq-3-3} \\
 V_{n|n} &=& (I -K_n H)V_{n|n-1}. \nonumber
\end{eqnarray}

Given the data $Y_N$, the likelihood of the time series model is defined by 
\begin{eqnarray}
 L(\theta ) &=& p( Y_N|\theta ) 
  =  \prod_{n=1}^N g_n(y_n|Y_{n-1},\theta ),
\end{eqnarray}
where $g_n(y_n|Y_{n-1},\theta )$ is the conditional distribution of $y_n$ given the observation $Y_{n-1}$ and is a normal distribution given by
\begin{eqnarray}
g_n(y_n|Y_{n-1},\theta ) 
  = \frac{1}{\sqrt{2\pi r_n}}\exp\left\{ -\frac{\varepsilon_{n}^2}{2r_n} \right\}, 
\label{eq_distribution_g}
\end{eqnarray}
where $\varepsilon_n$ and $r_n$ are the one-step-ahead prediction error and
its variance defined by
\begin{eqnarray}
\varepsilon_n &=& y_n - Hx_{n|n-1} \nonumber \\
r_n &=&  H_n V_{n|n-1} H_n^T + 1  \label{Eq_prediction_error}
\end{eqnarray}

Therefore, the log-likelihood of the state-space model is obtained as
\begin{eqnarray}
 \ell (\theta ) = \log L(\theta ) &=& \sum_{n=1}^N \log g_n(y_n|Y_{n-1},\theta ) \nonumber \\
  &=& -\frac{1}{2} \biggl( N \log 2\pi \hat\sigma^2 + \sum_{n=1}^N \log r_{n} +N
 \biggr), \label{Eq_log-lk}
\end{eqnarray}
where the maximum likelihood estimate of the variance is given by
\begin{eqnarray}
\hat\sigma^2 = \frac{1}{N}\sum_{n=1}^N \frac{\varepsilon_{n}^2}{r_n}.
\label{Eq_Observation_noise_variance}
\end{eqnarray}

The maximum likelihood estimates of the parameters of the state-space model 
can be obtained by maximizing the log-likelihood function (\ref{Eq_log-lk}).
In general, since the log-likelihood function is mostly nonlinear, the maximum likelihood estimates are obtained by using a numerical optimization algorithm based on the quasi-Newton method. 
According to this method, using the value  $\ell( \theta)$ of the log-likelihood and the first derivative (gradient) $\partial \ell/\partial \theta$ for a given parameter \( \theta \), the maximizer of \( \ell( \theta) \) is automatically estimated by repeating 
\begin{equation}
 \theta_k = \theta_{k-1} + \lambda_k B_{k-1}^{-1} \frac{\partial \ell}{\partial \theta},
\end{equation}
where $\theta_0$ is an initial estimate of the parameter. 
The step width \( \lambda_k \) is automatically determined and the inverse matrix \( H_{k-1}^{-1} \) of the Hessian matrix is obtained recursively by the DFP or BFGS algorithms (Fletcher (2013)).

Here, the gradient of the log-likelihood function is usually approximated by numerical difference,
such as 
\begin{eqnarray}
\frac{\partial \ell (\theta)}{\partial \theta_j} \approx \frac{\ell (\theta_j +\Delta \theta_j) -  \ell (\theta_j - \Delta \theta_j)}{2\Delta\theta_j},
\end{eqnarray}
where $\Delta\theta_j$ is defined by $C |\theta_j|$, for some small $C$ such as 0.0001.
The numerical difference usually yields reasonable approximation to the gradient of the 
log-likelihood.
However, since it requires $2p$ times of log-likelihood evaluations, the amount of computation
becomes considerablly large if the dimension of the parameters is large.
Further, if the the maximum likelihood estimates lie very close to the boundary of
addmissible domain, which sometimes occure in regularization problems,
it becomes difficult to obtain the approximation to the gradient of the log-likelihood
by  the numerical difference.

Analytic derivative of the log-likelihood of time series models were considered by many authors. 
For example,  Kohn and Ansley (1985) gave method for computing likelihood and its derivatives for an ARMA model.
Zadrozny (1989) derived analytic derivatives for estimation of linear dynamic models.
Kulikova (2009) presented square-root algorithm for the evaluation of the likelihood gradient to avoid numerical instability of the recursive algorithm for log-likelihood computation.

In this paper,  the gradient and Hessian of the log-likelihood of linear state-space model are given
under the assumption that the observation noise variance is 1. 
By this method, the dimension of the unknown parameter is  reduced by one, but instead 
the derivative of the observation noise variance must be computed simultaneously.
Details of the implementation of the algorithm for the trend model, the standard seasonal adjustment model and the seasonal adjustment model with stationary AR component are given.
For each implementation, comparison with a numerical difference method is shown.
In section 2, algorithm for obtaining the gradient and the Hessian of the log-likelihood
is presented.
Application of the method is exemplified with the three models, i.e., the trend model, the standard seasonal adjustment model and  the seasonal adjustment model
with autoregressive component are shown in section 3.


\section{The Gradient and the Hessian of the log-likelihood }

The recursive algorithm for computing the gradient and the Hessian of the 
log-likelihood is essentially the same as the one shown in Kitagawa(2020b).
However, we assume that the observation noise variance is 1, the expression 
for the gradient and the Hessain matrix become simple.
Instead, we need to evaluate the first and the second derivative of the
observation noise variance $\sigma^2$.

\subsection{The gradient of the log-likelihood}

From (\ref{Eq_log-lk}), the gradient of the log-likelihood is obtained by
\begin{eqnarray}
\frac{\partial\ell (\theta )}{\partial \theta} 
&=& - \frac{1}{2}\left( 
   \frac{N}{\hat\sigma^2}\frac{\partial \hat\sigma^2}{\partial\theta}
  + \sum_{n=1}^N  \frac{1}{r_n}\frac{\partial r_n}{\partial\theta}
   \right),  \label{Eq_gradient_ell}
\end{eqnarray}
where, from (\ref{Eq_prediction_error}) and (\ref{Eq_Observation_noise_variance}), the derivatives of the one-step-ahead predition $\varepsilon_n$, the
one-step-ahead prediction error variance $r_n$ and the observation
noise variance are obtained by
\begin{eqnarray}
\frac{\partial \varepsilon_n}{\partial\theta} &=& -H \frac{\partial x_{n|n-1}}{\partial\theta}
   - \frac{\partial H}{\partial\theta}x_{n|n-1}  \label{Eq_gradient_pred_error}\\
\frac{\partial r_n}{\partial\theta} &=& H \frac{\partial V_{n|n-1}}{\partial\theta}H^T
   + \frac{\partial H}{\partial\theta}V_{n|n-1}H^T 
   + H V_{n|n-1} \frac{\partial H}{\partial\theta}^T
   + \frac{\partial R}{\partial\theta} , \label{Eq_gradient_observation_error}\\
\frac{\partial \hat\sigma^2}{\partial\theta} &=& \frac{1}{N}\sum_{n=1}^N  \left( \frac{2\varepsilon_n}{r_n} \frac{ \partial\varepsilon_{n}}{\partial\theta} 
 -  \frac{\varepsilon_{n}^2}{r_n^2} \frac{\partial r_n}{\partial\theta}  \right).
\label{Eq_gradient_observation_noise_variance}
\end{eqnarray}

To evaluate these quantities, we need the derivatives of the one-step-ahead predictor
of the state $\displaystyle\frac{\partial x_{n|n-1}}{\partial\theta}$ and its variance covariance
matrix $\displaystyle\frac{\partial V_{n|n-1}}{\partial\theta}$ which can be obtained recursively
in parallel to the Kalman filter algorithm:\\

\quad [One-step-ahead-prediction] \\
\begin{eqnarray}
\frac{\partial x_{n|n-1}}{\partial\theta} &=& F \frac{\partial x_{n-1|n-1}}{\partial\theta}
   + \frac{\partial F}{\partial\theta} x_{n-1|n-1}
\nonumber \\
\frac{\partial V_{n|n-1}}{\partial\theta} &=& F \frac{\partial V_{n-1|n-1}}{\partial\theta}F^T 
           + \frac{\partial F}{\partial\theta}V_{n-1|n-1}F^T 
           + F V_{n-1|n-1} \frac{\partial F}{\partial\theta}^T \nonumber \\
         &&  + G\frac{\partial Q}{\partial\theta}G^T 
           + \frac{\partial G}{\partial\theta}Q G^T 
           + G Q \frac{\partial G}{\partial\theta}^T .  \label{Eq_gradient-filter-P}
\end{eqnarray}

\quad [Filter] \\
\begin{eqnarray}
\frac{\partial K_n}{\partial\theta} 
        &=& \left(\frac{\partial V_{n|n-1}}{\partial\theta}H^T
               +   V_{n|n-1}\frac{\partial H}{\partial\theta}^T\right) r_n^{-1} 
          - V_{n|n-1}H^T \frac{\partial r_n}{\partial\theta}r_n^{-2 }    \nonumber \\
\frac{\partial x_{n|n}}{\partial\theta} &=&  
        \frac{\partial x_{n|n-1}}{\partial\theta}
       + K_n \frac{\partial \varepsilon_n}{\partial\theta}
       + \frac{\partial K_n}{\partial\theta} \varepsilon_n   \nonumber \\
\frac{\partial V_{n|n}}{\partial\theta} &=& 
         \frac{\partial V_{n|n-1}}{\partial\theta}
       - \frac{\partial K_n}{\partial\theta} H V_{n|n-1}
       - K_n \frac{\partial H}{\partial\theta} V_{n|n-1}
       - K_n H \frac{\partial V_{n|n-1}}{\partial\theta}. \label{Eq_gradient-filter-F}
\end{eqnarray}


\subsection{Hessian of the Log-likelihood of the State-space Model}

The Hessian (the second derivative) of the log-likelihood can 
also be obtained by a recursive formula, 
since, from (\ref{Eq_gradient_ell}), it is given as
\begin{eqnarray}
\frac{\partial^2\ell (\theta )}{\partial \theta\partial \theta^T}
= - \frac{N}{2} \left( 
    \frac{1}{\hat\sigma^2}\frac{\partial^2 \hat\sigma^2}{\partial\theta\partial\theta^T} 
  - \frac{1}{(\hat\sigma^2)^2} \frac{\partial \hat\sigma^2}{\partial\theta}\frac{\partial \hat\sigma^2}{\partial\theta^T}  \right)   
- \frac{1}{2}\sum_{n=1}^N  \left( 
 \frac{1}{r_n}\frac{\partial^2 r_n}{\partial\theta \partial\theta^T} 
- \frac{1}{r_n^2}\frac{\partial r_n}{\partial\theta }\frac{\partial r_n}{\partial\theta^T}  \right)
, \nonumber
\end{eqnarray}
where, from (\ref{Eq_gradient_observation_error})--(\ref{Eq_gradient_observation_noise_variance}), 
$ \displaystyle\frac{\partial^2 \varepsilon_n}{\partial\theta\partial\theta^T} $, 
$ \displaystyle\frac{\partial^2 r_n}{\partial\theta \partial\theta^T} $
and $ \displaystyle\frac{\partial^2 \hat\sigma^2_n}{\partial\theta \partial\theta^T} $ are
obtained by
\begin{eqnarray}
\frac{\partial^2 \varepsilon_n}{\partial\theta\partial\theta^T} 
&=&  -  \frac{\partial H}{\partial\theta^T}\frac{\partial x_{n|n-1}}{\partial\theta}
-  \frac{\partial H}{\partial\theta}\frac{\partial x_{n|n-1}}{\partial\theta^T} 
-  H\frac{\partial^2 x_{n|n-1}}{\partial\theta\partial\theta^T}
-  \frac{\partial^2 H}{\partial\theta \partial\theta^T}x_{n|n-1} 
  \\
\frac{\partial^2 r_n}{\partial\theta \partial\theta^T} 
   &=& \frac{\partial H}{\partial\theta^T}\frac{\partial V_{n|n-1}}{\partial\theta}H^T
     + \frac{\partial H}{\partial\theta}\frac{\partial V_{n|n-1}}{\partial\theta^T}H^T 
     + H \frac{\partial^2 V_{n|n-1}}{\partial\theta \partial\theta^T}H^T  \nonumber \\
   &&+ H \frac{\partial V_{n|n-1}}{\partial\theta^T}\frac{\partial H}{\partial\theta} 
     + H \frac{\partial V_{n|n-1}}{\partial\theta}\frac{\partial H}{\partial\theta^T}   
     + \frac{\partial^2 H}{\partial\theta\partial\theta^T}V_{n|n-1}H^T \\
   &&+ \frac{\partial H}{\partial\theta^T}V_{n|n-1}\frac{\partial H^T}{\partial\theta} 
     + \frac{\partial H}{\partial\theta}V_{n|n-1}\frac{\partial H^T}{\partial\theta^T} 
     + H V_{n|n-1} \frac{\partial^2 H}{\partial\theta \partial\theta^T}
     + \frac{\partial^2 R}{\partial\theta\partial\theta^T}. \nonumber \\
\frac{\partial^2 \hat\sigma^2}{\partial\theta \partial\theta^T} 
&=& \frac{1}{N}\sum_{n=1}^N  \left\{ 
      \frac{2}{r_n}\!\!\left( \frac{ \partial \varepsilon_n}{\partial\theta} \frac{ \partial \varepsilon_n}{\partial\theta^T}  
   +  \varepsilon_n \frac{\partial^2 \varepsilon_n}{\partial\theta \partial\theta^T} \right)
   +  \frac{2\varepsilon_n^2}{r_n^3}  \frac{\partial r_n}{\partial\theta} \frac{\partial r_n}{\partial\theta^T} 
 \right. \nonumber \\
&&{}\qquad\qquad  \left. -  \frac{1}{r_n^2}\left(   
      2\varepsilon_n \frac{\partial \varepsilon_n}{\partial\theta^T} \frac{\partial r_n}{\partial\theta}   
   +  2\varepsilon_n \frac{\partial \varepsilon_n}{\partial\theta} \frac{\partial r_n}{\partial\theta^T} 
   +  \varepsilon_n^2 \frac{\partial^2 r_n}{\partial\theta \partial\theta^T} \right) \right\}
\end{eqnarray}

To evaluate the Hessian, the following computation should be performed
along with the recursive formula for the log-likelihood and the 
gradient of the log-likelihood.%

\begin{eqnarray}
\frac{\partial^2 x_{n|n-1}}{\partial\theta \partial\theta^T}
   &=&  \frac{\partial F}{\partial\theta^T} \frac{\partial x_{n-1|n-1}}{\partial\theta}
     + \frac{\partial F}{\partial\theta} \frac{\partial x_{n-1|n-1}}{\partial\theta}
     + F \frac{\partial^2 x_{n-1|n-1}}{\partial\theta \partial\theta^T}
     + \frac{\partial^2 F}{\partial\theta \partial\theta^T} x_{n-1|n-1}
\nonumber \\
\frac{\partial^2 V_{n|n-1}}{\partial\theta \partial\theta^T}
         &=& \frac{\partial F}{\partial\theta^T} \frac{\partial V_{n-1|n-1}}{\partial\theta}F^T  
           + \frac{\partial F}{\partial\theta} \frac{\partial V_{n-1|n-1}}{\partial\theta^T}F^T  
         + F \frac{\partial^2 V_{n-1|n-1}}{\partial\theta \partial\theta^T}F^T  
         + F \frac{\partial V_{n-1|n-1}}{\partial\theta^T}\frac{\partial F^T}{\partial\theta}  \nonumber \\
       && + F \frac{\partial V_{n-1|n-1}}{\partial\theta}\frac{\partial F^T}{\partial\theta^T} 
          + \frac{\partial^2 F}{\partial\theta \partial\theta^T}V_{n|n-1}F^T 
          + \frac{\partial F}{\partial\theta^T}V_{n|n-1}\frac{\partial F^T}{\partial\theta} 
          + \frac{\partial F}{\partial\theta}V_{n|n-1}  \frac{\partial F^T}{\partial\theta^T}  \nonumber \\
       && + F V_{n|n-1} \frac{\partial^2 F^T}{\partial\theta \partial\theta^T} 
          + \frac{\partial G}{\partial\theta^T} \frac{\partial Q}{\partial\theta}G^T 
          + \frac{\partial G}{\partial\theta} \frac{\partial Q}{\partial\theta^T}G^T 
          + G\frac{\partial^2 Q}{\partial\theta \partial\theta^T}G^T 
          + G\frac{\partial Q}{\partial\theta^T}\frac{\partial G^T}{\partial\theta} \nonumber \\
       && +  G\frac{\partial Q}{\partial\theta}\frac{\partial G^T}{\partial\theta^T}  
          + \frac{\partial^2 G}{\partial\theta\partial\theta^T}Q G^T 
          + \frac{\partial G}{\partial\theta^T}Q \frac{\partial G^T}{\partial\theta} 
          + \frac{\partial G}{\partial\theta}Q \frac{\partial G^T}{\partial\theta^T} 
          + G Q \frac{\partial^2 G^T}{\partial\theta \partial\theta^T} \nonumber \\
\frac{\partial^2 K_n}{\partial\theta \partial\theta^T} 
        &=& \left(\frac{\partial^2 V_{n|n-1}}{\partial\theta \partial\theta^T}H^T  
             + \frac{\partial V_{n|n-1}}{\partial\theta^T}\frac{\partial H}{\partial\theta}
             + \frac{\partial V_{n|n-1}}{\partial\theta}\frac{\partial H}{\partial\theta}^T 
             + V_{n|n-1}\frac{\partial^2 H}{\partial\theta \partial\theta^T}
           \right) r_n^{-1} \nonumber \\
        &&  - \left( \frac{\partial V_{n|n-1}}{\partial\theta^T}H^T  
              \frac{\partial r_n}{\partial\theta}
             + \frac{\partial V_{n|n-1}}{\partial\theta}H^T  \frac{\partial r_n}{\partial\theta^T}
             + V_{n|n-1}\frac{\partial H}{\partial\theta^T}^T \frac{\partial r_n}{\partial\theta}                          + V_{n|n-1}\frac{\partial H}{\partial\theta}^T \frac{\partial r_n}{\partial\theta^T}   \right.
        \nonumber \\
       && {}\quad \left. +  V_{n|n-1}H^T  \frac{\partial^2 r_n}{\partial\theta \partial\theta^T}  
           \right) r_n^{-2}  
          + 2V_{n|n-1}H^T 
            \frac{\partial r_n}{\partial\theta} \frac{\partial r_n}{\partial\theta^T} r_n^{-3} 
    \label{Eq_Hessian_filter}\\
\frac{\partial^2 x_{n|n}}{\partial\theta \partial\theta^T} &=&  
        \frac{\partial^2 x_{n|n-1}}{\partial\theta \partial\theta^T}
       +  \frac{\partial K_n}{\partial\theta} \frac{\partial \varepsilon_n}{\partial\theta^T}
       + \frac{\partial K_n}{\partial\theta^T} \frac{\partial \varepsilon_n}{\partial\theta}  
       + K_n \frac{\partial^2 \varepsilon_n}{\partial\theta \partial\theta^T} 
       + \frac{\partial^2 K_n}{\partial\theta \partial\theta^T} \varepsilon_n
\nonumber \\
\frac{\partial^2 V_{n|n}}{\partial\theta \partial\theta^T} &=& 
         \frac{\partial^2 V_{n|n-1}}{\partial\theta \partial\theta^T}
       - \frac{\partial^2 K_n}{\partial\theta \partial\theta^T} H V_{n|n-1}
       - \frac{\partial K_n}{\partial\theta^T} \frac{\partial H}{\partial\theta} V_{n|n-1}
       - \frac{\partial K_n}{\partial\theta} \frac{\partial H}{\partial\theta^T} V_{n|n-1}  \nonumber \\
    && - \frac{\partial K_n}{\partial\theta^T} H \frac{\partial V_{n|n-1}}{\partial\theta} 
       - K_n \frac{\partial^2 H}{\partial\theta \partial\theta^T} V_{n|n-1}
       - \frac{\partial K_n}{\partial\theta} H \frac{\partial V_{n|n-1}}{\partial\theta^T}   
\nonumber \\
    &&
       -   K_n \frac{\partial H}{\partial\theta} \frac{\partial V_{n|n-1}}{\partial\theta^T} 
       - K_n \frac{\partial H}{\partial\theta^T} \frac{\partial V_{n|n-1}}{\partial\theta}  
       - K_n H \frac{\partial^2 V_{n|n-1}}{\partial\theta \partial\theta^T}. \nonumber
\end{eqnarray}

%

\section{Examples}
In order to implement the differential filter, it is necessary to to specify 
the first and the second derivatives
of $F$, $G$, $H$ and $Q$ along with the original state-space model.
In this section, we shall consider three typical cases.
The first two examples are the trend model and the standard seasonal adjeustment model, 
for which three matrices (or vectors), $F$, $G$ and $H$ do not contain unknown 
parameters and thus the derivatives
of these matrics becomes 0. 
This makes the algorithm for the gradient and the Hessian of the log-likelihood 
presented in the previous section considerablly simple.
The third example is the seasonal adjustment model with AR component.
For this model,  the matrix $F$ depends on the unknown AR coefficients, 
although the derivative of $F$ is very sparse.
However, since we usually use a nonlinear transformation of the parameters 
and the Levinson's formula between partial autocorrelations coefficients
and AR coefficients, to ensure the stationarity condition,
the expression of the non-zero elements of the derivatives of $F$ becomes fairly complex.

\subsection{Trend model}

The trend model is a typical example of the case where only the noise covariance $Q$ 
depends on the unknown parameter $\theta$.
Consider a trend model
\begin{eqnarray}
y_n = T_n + w_n,
\end{eqnarray}
where $T_n$ is the trend component 
that typically follow the following model
\begin{eqnarray}
(1-B)^k T_n = v_n, \label{eq_trend_model}
\end{eqnarray}
where $B$ is the back-shift operator satisfying $BT_n = T_{n-1}$,
 $v_n$ and $w_n$ are assumed to be Gaussian white noise with
variances $\tau^2$ and $1$, respectively (Kitagawa and Gersch (1984,1996) and 
Kitagawa (2020a)).
Note that for $k=1$ and $k=2$, the model (\ref{eq_trend_model}) becomes
$T_n = T_{n-1}+v_n$ and $T_n = 2T_{n-1} - T_{n-2} + v_n$, respectively,
and that in the Kalman filter, the essentially the same filtering 
results can be obtained by assuming that $\sigma^2 =1$ (Kitagawa (2020a))
and thus the dimension of the unknown parameter vector is reduced by one,
i.e., in the case of the trend model the dimension of the parameter
becomes one.

This trend model can be expressed in
the state-space model form as
\begin{eqnarray}
x_n &=& F x_{n-1} + G v_n \nonumber \\
y_n &=& H x_n + w_n,
\end{eqnarray}
with $v_n \sim N(0,Q)$ and $w_n \sim N(0,1)$ and the state vector $x_n$ and 
the matrices $F$, $G$, $H$, $Q$ and $R$ are defined by
\begin{eqnarray}
x_n &=& T_n,\quad  F = 1 ,\quad G = 1, \quad H =1, \quad
Q = \tau^2, \quad  R = 1,
\end{eqnarray}  
for $k=1$ and
\begin{eqnarray}
x_n &=& \left[ \begin{array}{c} 
             T_n     \\
             T_{n-1}  \end{array}\right],\quad
F = \left[ \begin{array}{cc} 
             2 &-1  \\
             1 & 0  \end{array}\right],\quad
G = \left[ \begin{array}{cc} 
             1 & 0   \\
             0 & 0   \end{array}\right] \\
H &=& [\begin{array}{cc} 1 &0 \end{array}], \quad
Q = \tau^2, \quad  R = 1,
\end{eqnarray}  
for $k=2$.

In this state-space representation, the parameter is $\tau^2$,
and the $F$, $G$, $H$ and $R$ do not depend on the parameter.
Therefore, we have $\displaystyle\frac{\partial F}{\partial\theta}=
\displaystyle\frac{\partial G}{\partial\theta}=\displaystyle\frac{\partial H}{\partial\theta}
=\displaystyle\frac{\partial R}{\partial\theta}=0$
and $\displaystyle\frac{\partial^2 F}{\partial\theta\partial\theta^T}=
\displaystyle\frac{\partial^2 G}{\partial\theta\partial\theta^T}=
\displaystyle\frac{\partial^2 H}{\partial\theta\partial\theta^T}=
\displaystyle\frac{\partial^2 R}{\partial\theta\partial\theta^T}=0$.

\vspace{1mm}
In actual likelihood maximization, since there is a positivity 
constraint, $\tau^2 > 0$,
it is frequently used a log-transformation,
\begin{eqnarray}
 \theta = \log (\tau^2) ,
\end{eqnarray}
and maximize the log-likelihood with respect to this transformed parameter $\theta$.
In this case, 
\begin{eqnarray}
  \frac{\partial Q}{\partial\theta} = 
  \frac{\partial^2 Q}{\partial\theta \partial\theta} = \tau^2. \quad
\end{eqnarray}
Since log-transfomation is a monotone incresing function, we can get the 
same parameter by solving this modified optimization problem.

In this case, the recursive algorithm for gradient of the log-likelihood 
becomes significantly simple as follows:

\begin{eqnarray}
\frac{\partial\ell (\theta )}{\partial \theta} 
&=& - \frac{1}{2}\left( 
   \frac{N}{\hat\sigma^2}\frac{\partial \hat\sigma^2}{\partial\theta}
  + \sum_{n=1}^N  \frac{1}{r_n}\frac{\partial r_n}{\partial\theta}
   \right),  \label{Eq_gradient_ell_trend} 
\end{eqnarray}
where, from (\ref{Eq_prediction_error}), the derivatives of the one-step-ahead predition error $\varepsilon_n$, the
one-step-ahead prediction error variance $r_n$ and the observation noise variance
$\hat\sigma^2$ are obtained by
\begin{eqnarray}
\frac{\partial \varepsilon_n}{\partial\theta_i} &=& -H \frac{\partial x_{n|n-1}}{\partial\theta_i}
    \nonumber \\
\frac{\partial r_n}{\partial\theta_i} &=& H \frac{\partial V_{n|n-1}}{\partial\theta_i}H^T  \\
\frac{\partial \hat\sigma^2}{\partial\theta} &=& \frac{1}{N}\sum_{n=1}^N  \left( \frac{2\varepsilon_n}{r_n} \frac{ \partial\varepsilon_{n}}{\partial\theta} 
 -  \frac{\varepsilon_{n}^2}{r_n^2} \frac{\partial r_n}{\partial\theta}  \right). 
\end{eqnarray}

The formula for obtaining the derivatives of the state and its variance covariance matrix become
\begin{eqnarray}
%
\frac{\partial x_{n|n-1}}{\partial\theta_i} &=& F \frac{\partial x_{n-1|n-1}}{\partial\theta_i}
\nonumber \\
\frac{\partial V_{n|n-1}}{\partial\theta_i} &=& F \frac{\partial V_{n-1|n-1}}{\partial\theta_i}F^T 
           + G\frac{\partial Q}{\partial\theta_i}G^T \nonumber \\
\frac{\partial K_n}{\partial\theta_i} 
        &=& \frac{\partial V_{n|n-1}}{\partial\theta_i}H^T r_n^{-1}  
      - V_{n|n-1}H^Tr_n^{-2} \frac{\partial r_n}{\partial\theta_i}  \nonumber  \\
\frac{\partial x_{n|n}}{\partial\theta_i} &=&  
       \frac{\partial x_{n|n-1}}{\partial\theta_i} 
       + \frac{\partial K_n}{\partial\theta_i} \varepsilon_n
       + K_n\frac{\partial \varepsilon_n}{\partial\theta_i}   \nonumber \\
\frac{\partial V_{n|n}}{\partial\theta_i} &=& 
         (I- K_nH)\frac{\partial V_{n|n-1}}{\partial\theta_i}
       - \frac{\partial K_n}{\partial\theta_i} H V_{n|n-1}. \nonumber
\end{eqnarray}

The Hessian (the second derivative) of the log-likelihood is also obtained
by a recursive formula, since, from (\ref{Eq_gradient_ell}), it is given as
\begin{eqnarray}
\frac{\partial^2\ell (\theta )}{\partial \theta\partial \theta^T}
&=& - \frac{N}{2} \left( 
    \frac{1}{\hat\sigma^2}\frac{\partial^2 \hat\sigma^2}{\partial\theta\partial\theta^T} 
  - \frac{1}{(\hat\sigma^2)^2} \frac{\partial \hat\sigma^2}{\partial\theta}\frac{\partial \hat\sigma^2}{\partial\theta^T}  \right)   
+  \frac{1}{2}\sum_{n=1}^N  \left( 
 \frac{1}{r_n}\frac{\partial^2 r_n}{\partial\theta \partial\theta^T} 
- \frac{1}{r_n^2}\frac{\partial r_n}{\partial\theta }\frac{\partial r_n}{\partial\theta^T}  \right)
, \nonumber 
\end{eqnarray}
where, from (\ref{Eq_gradient_observation_error}), $ \displaystyle\frac{\partial^2 \varepsilon_n}{\partial\theta_i\partial\theta_j} $,
$ \displaystyle\frac{\partial^2 r_n}{\partial\theta_i \partial\theta_j} $ 
and $ \displaystyle\frac{\partial^2 \hat\sigma^2}{\partial\theta_i \partial\theta_j} $ 
are obtained by
\begin{eqnarray}
\frac{\partial^2 \varepsilon_n}{\partial\theta_i\partial\theta_j} 
&=&  -  H\frac{\partial^2 x_{n|n-1}}{\partial\theta_i \partial\theta_j}
 \nonumber \\
\frac{\partial^2 r_n}{\partial\theta_i \partial\theta_j} 
   &=&  H \frac{\partial^2 V_{n|n-1}}{\partial\theta_i \partial\theta_j}H^T  \\
\frac{\partial^2 \hat\sigma^2}{\partial\theta \partial\theta^T} 
&=& \frac{1}{N}\sum_{n=1}^N  \left\{ 
      \frac{2}{r_n}\!\!\left( \frac{ \partial \varepsilon_n}{\partial\theta} \frac{ \partial \varepsilon_n}{\partial\theta^T}  
   +  \varepsilon_n \frac{\partial^2 \varepsilon_n}{\partial\theta \partial\theta^T} \right)
   +  \frac{2\varepsilon_n^2}{r_n^3}  \frac{\partial r_n}{\partial\theta} \frac{\partial r_n}{\partial\theta^T} 
 \right. \nonumber \\
 &&  {}\qquad\quad  -  \frac{1}{r_n^2} \left. \left(  
      2\varepsilon_n \frac{\partial \varepsilon_n}{\partial\theta^T} \frac{\partial r_n}{\partial\theta}  
   +  2\varepsilon_n \frac{\partial \varepsilon_n}{\partial\theta} \frac{\partial r_n}{\partial\theta^T}  
   +  \varepsilon_n^2 \frac{\partial^2 r_n}{\partial\theta \partial\theta^T} \right) \right\}
\end{eqnarray}

Therefore, to evaluate the Hessian, the following computation should be performed
along with the recursive formula for the log-likelihood and the 
gradient and Hessian of the log-likelihood. 
Note that since in this example, the parameter is one dimentional,
we should read $i=j$ in the following formula.

\begin{eqnarray}
\frac{\partial^2 x_{n|n-1}}{\partial\theta_i \partial\theta_j}
   &=& F \frac{\partial^2 x_{n-1|n-1}}{\partial\theta_i \partial\theta_j}
\nonumber \\
\frac{\partial^2 V_{n|n-1}}{\partial\theta_i \partial\theta_j}
         &=& F \frac{\partial^2 V_{n-1|n-1}}{\partial\theta_i \partial\theta_j}F^T  
        + G\frac{\partial^2 Q}{\partial\theta_i \partial\theta_j}G^T   \nonumber \\ %
\frac{\partial^2 K_n}{\partial\theta_i \partial\theta_j} 
        &=& r_n^{-1} \frac{\partial^2 V_{n|n-1}}{\partial\theta_i \partial\theta_j}H^T  
         + 2 r_n^{-3} V_{n|n-1}H^T 
           \frac{\partial r_n}{\partial\theta_i} \frac{\partial r_n}{\partial\theta_j} 
                     \nonumber \\
       && -  r_n^{-2}\left(  \frac{\partial V_{n|n-1}}{\partial\theta_i}H^T  
             \frac{\partial r_n}{\partial\theta_j}  
         +  \frac{\partial V_{n|n-1}}{\partial\theta_j}H^T  
             \frac{\partial r_n}{\partial\theta_i}
         +  V_{n|n-1}H^T \frac{\partial^2 r_n}{\partial\theta_i \partial\theta_j}     \right)   
   \\
\frac{\partial^2 x_{n|n}}{\partial\theta_i \partial\theta_j} &=&  
        \frac{\partial^2 x_{n|n-1}}{\partial_i\theta \partial\theta_j}
        + K_n \frac{\partial^2 \varepsilon_n}{\partial\theta_i \partial\theta_j} 
       + \frac{\partial K_n}{\partial\theta_i} \frac{\partial \varepsilon_n}{\partial\theta_j}
       + \frac{\partial K_n}{\partial\theta_j} \frac{\partial \varepsilon_n}{\partial\theta_i}
       + \frac{\partial^2 K_n}{\partial\theta_i \partial\theta_j} \varepsilon_n
\nonumber \\
\frac{\partial^2 V_{n|n}}{\partial\theta_i \partial\theta_j} &=& 
         (I -K_n H) \frac{\partial^2 V_{n|n-1}}{\partial\theta_i \partial\theta_j}
       - \frac{\partial^2 K_n}{\partial\theta_i \partial\theta_j} H V_{n|n-1}
       - \frac{\partial K_n}{\partial\theta_i} H \frac{\partial V_{n|n-1}}{\partial\theta_j} 
       - \frac{\partial K_n}{\partial\theta_j} H \frac{\partial V_{n|n-1}}{\partial\theta_i} 
. \nonumber
\end{eqnarray}

\begin{table}[bp]
\begin{center}
\caption{Comparison of the results by the numerical diffference and the proposed analytic method for the first order trend model ($m_1=1$). Left: initial values, Right: final estimate.}\label{Tab_Trend_model_m1=1}

\vspace{2mm}
\begin{tabular}{c|cc|cc}
        &  Difference &  Analytic &  Difference &  Analytic \\
\hline 
  $ \tau^2$ &  0.50000      & 0.50000     & 4.49933 & 4.49933 \\
  $ \theta$ &  $-0.69315$   & $-0.69315$  & 1.50393 & 1.50393 \\ 
  $ \ell(\theta )$ & $-307.6616 $& $-307.6616$ & 317.9243 &  317.9243 \\[2mm]
  $\frac{\partial \ell(\theta )}{\partial \theta}$ & 9.87647 & 9.87647  & $-2.7365\!\times \!\!10^{-7}$   &$-2.86304\!\times \!\!10^{-7}$   \\[2mm]
  $\frac{\partial^2 \ell(\theta )}{\partial \theta\partial \theta}$ & $-3.54544$ & $-3.54545$  &  $-2.34221$ & $-2.34219$ \\[2mm]
 $\hat\sigma^2$  & $5.50118\!\times \!\!10^{-4}$ & $5.50152\!\times \!\!10^{-4}$ & $1.52463\!\times \!\!10^{-4}$   &$1.52497\!\times \!\!10^{-4}$ \\
\hline
\end{tabular}
\end{center}
\end{table}

For Whard (whole sale hardware) data (Kitagawa (2020a)), $N=155$, 
the system noise variance parameter $\theta =\tau^2$ of the trend model 
with $m_1=1$ was estimated using the initial value $\tau^2 = 0.5$, i.e., $\theta_0 = \log (0.5) = -0.69315$. 
By a numerical optimization procedure the maximum likelihood estimate of the parameter is obtained
 as  $\hat\theta = 1.50393$, i.e., $ \hat\tau^2 = e^{\hat\theta} = 4.49933 $.

Table \ref{Tab_Trend_model_m1=1} shows the log-likelihoods, the gradients, the Hessians and the 
observation noise variances of the initial and the final estimates.
For comparison, the values  obtained by the numerical difference are also shown in the table.
The log-likelihood of the model with these initial and final estimates are 
$\ell (\theta )= -307.6616$ and $-317.9243$, respectively.
It can be seen that the analytic gradient and the Hessian coincide with ones obtained by the numerical differentiation at least up to 4th digit.

\begin{figure}[tbp]
\begin{center}
\includegraphics[width=150mm,angle=0,clip=]{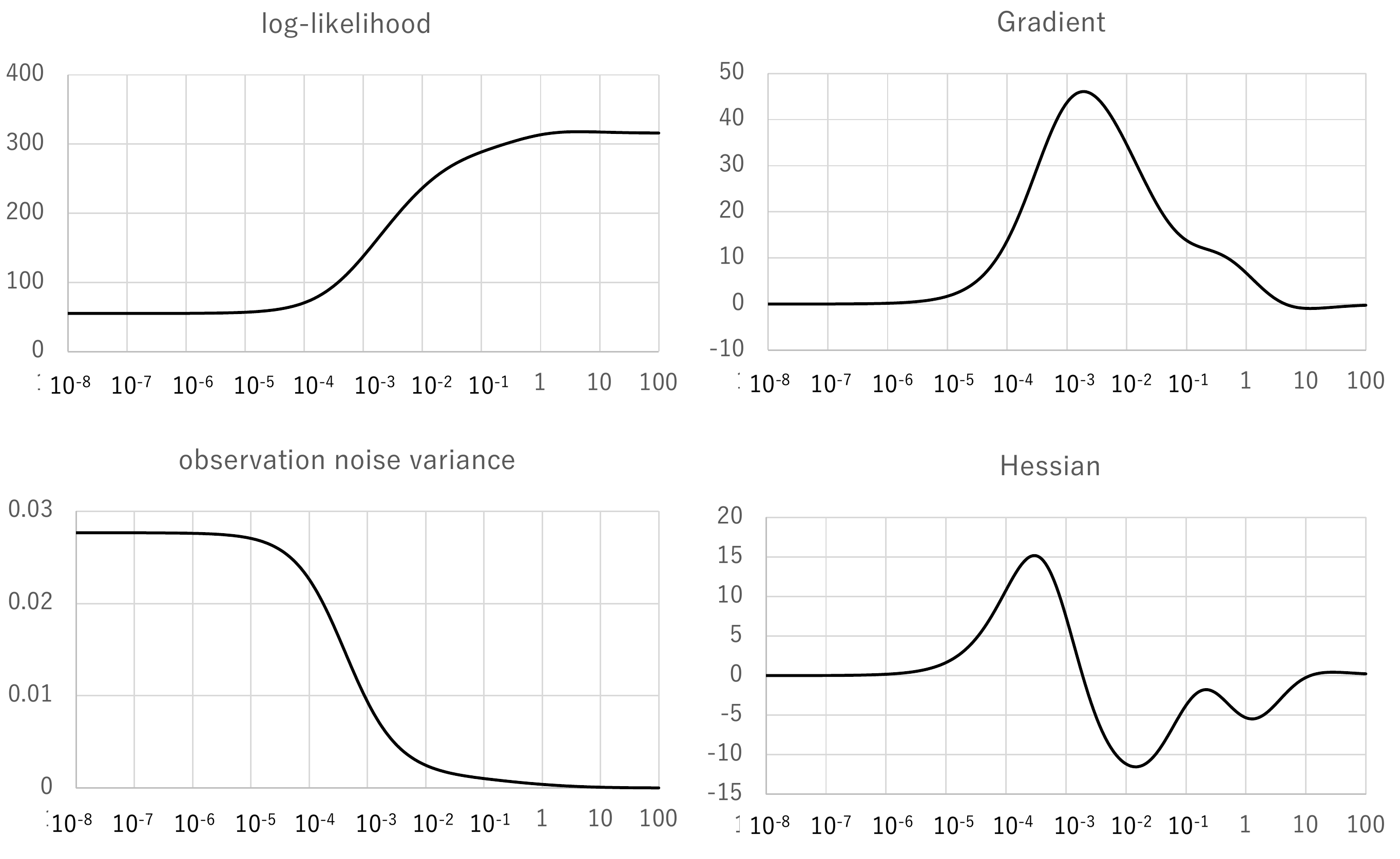}
\end{center}
\caption{The log-likelihood, the gradient, the Hessian and the variance of the observation
noise of the trend models with order 1. The horizontal axes are the value of the system noise
variance in log-scale. }
\label{Fig_Trend model with order 1}
\vspace{5mm}
\end{figure}

Figure \ref{Fig_Trend model with order 1} shows the change of the log-likelihood, the gradient,
Hessian and the observation noise variance for various values of the system noise vaiance
$\tau^2$.
In this case, the log-likelihood has only one local maximum and the gradient is unimodal.
On the other hand, the Hessian has two peakes and two troughs.

\begin{table}[tbp]
\begin{center}
\caption{Comparison of the results by the numerical diffference and the proposed analytic method for the second order trend model ($m_1=2$)}\label{Tab_Trend_model_m1=2}

\vspace{2mm}
\begin{tabular}{c|cc|cc}
        &  Difference &  Analytic &  Difference &  Analytic \\
\hline
  $\theta_0$ & $-13.8155$ & $-13.8155$ & -0.53318  & -0.53318\\
  $\tau^2_0$ & $0.100\times \!10^{-5}$ & $0.100\times \!10^{-5}$ & 0.58674  & 0.58674 \\
  $ \ell(\theta )$ & -270.02065 & -270.02065 &  296.17899 &   296.17899 \\[2mm]
  $\frac{\partial \ell(\theta )}{\partial \theta}$ & 1.68866 & 1.68866  & $5.6238\!\times \!\!10^{-7}$   &$5.5944\!\times \!\!10^{-7}$   \\[2mm]
  $\frac{\partial^2 \ell(\theta )}{\partial \theta\partial \theta}$ & 0.010951 & 0.010953 &  -6.150639 & -6.15072 \\[2mm]
 $\hat\sigma^2$  & $1.57779\!\times \!\!10^{-3}$ & $1.57791\!\times \!\!10^{-3}$ & $3.55274\!\times \!\!10^{-4}$  &$3.55286\!\times \!\!10^{-4}$ \\
\hline
  $\hat\theta$ & $-6.90776$ & $-6.90776$ & $-7.0399$  & $-7.0399$ \\
  $\hat\tau^2$ & $0.1000\times \!10^{-2}$ & $0.1000\times \!10^{-2}$ & $0.87619\times \!10^{-3}$  & $0.87619\times \!10^{-3}$ \\
  $ \ell(\theta )$ & -283.72502 & -283.72502 &  283.73930 &   283.73930 \\[2mm]
  $\frac{\partial \ell(\theta )}{\partial \theta}$ & -0.214378 & -0.21437  & $0.24466\times 10^{-8}$  &$-0.6156\times 10^{-8}$   \\[2mm]
  $\frac{\partial^2 \ell(\theta )}{\partial \theta\partial \theta}$ & -1.58184 & -1.58183  & -1.66045  & -1.66045 \\[2mm]
 $\hat\sigma^2$  & $1.1221\times \!10^{-3}$ & $1.12215\times \!10^{-3}$ &  $1.13024\times \!10^{-3}$ &$1.13029\times \!10^{-3}$ \\
\hline
\end{tabular}
\end{center}
\end{table}

\begin{figure}[tbp]
\begin{center}
\includegraphics[width=140mm,angle=0,clip=]{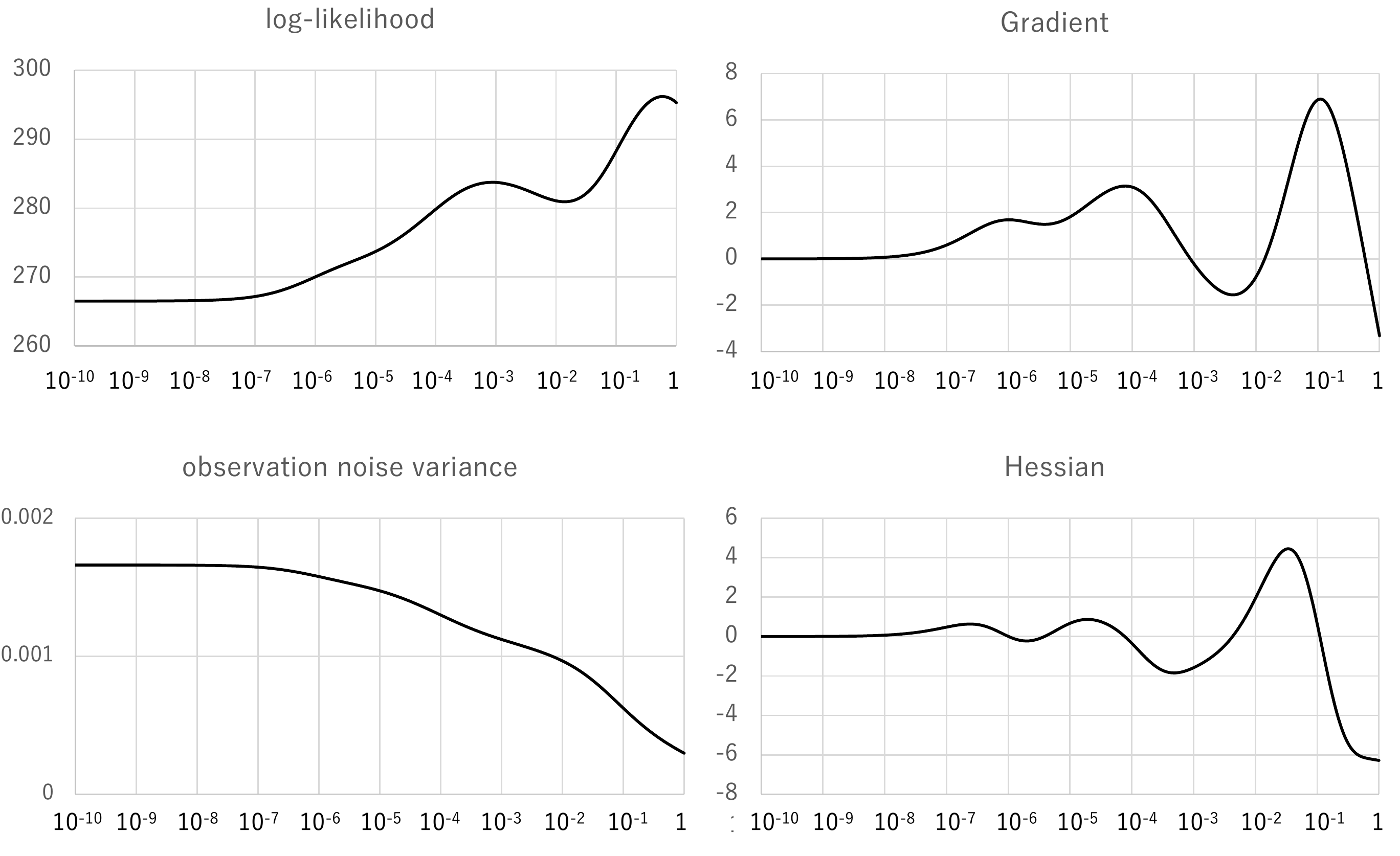}
\end{center}
\caption{The log-likelihood, the gradient, the Hessian and the variance of the observation
noise of the trend models with order 2. The horizontal axes are the value of the system noise
variance in log-scale. }
\label{Fig_Trend model with order 2}
\vspace{5mm}
\end{figure}

Table \ref{Tab_Trend_model_m1=2} shows the results for the sencond order trend model.
In this case, the final estimate obtained by the numerical optimization procedure
depends on the initial estimate and two cases are shown in the table.
If the inital estiamte is set to $\theta_0=-13.8155$, i.e., $\tau^2_0 =10^{-4}$, the final estimate is $\hat\theta=-0.53318$, i.e., $\hat\tau^2_0=0.58674$
with the log-likelihood value $\ell (\hat\theta )=296.179$.
On the other hand, if we set the initial estimate as $\theta_0=-6.90776$, i.e., $\tau^2_0=10^{-1}$, the final
estiamte becomes $\hat\theta = -7.0399$, i.e., $\hat\tau^2 =0.87619\times 10^{-3}$ with $\ell (\hat\theta )=283.739$.
Comparing the log-likelihodd values, $\hat\tau^2 = 0.58674$ is the maximum likelihood 
estimate of the second order trend model.

Figure \ref{Fig_Trend model with order 2} shows the change of the log-likelihood, the gradient,
the Hessian and the observation noise variance for various values of the system noise vaiance
$\tau^2$ for the second order trend model.
In this case, the log-likelihood is bimodal and the gradient attains zero at three points, 
two local maxima and one local minimum.
The Hessian of the two local maximum likelihood estiamte, 0.58674 and $0.87619\times 10^{-3}$
are 6.15072 and 1.66045, respectively.
This indicates that the estimate 0.58674 has sharper peak in the log-likelihood function.



\newpage

\subsection{The standard seasonal adjustment model}

As the second example, we consider a standard seasonal adjustment model
\begin{eqnarray}
y_n = T_n + S_n + w_n,
\end{eqnarray}
where $T_n$ and $S_n$ are the trend component and the seasonal component
that typically follow the following model
\begin{eqnarray}
&&T_n = 2T_{n-1} - T_{n-2} + u_n, \nonumber \\
&&S_n =-(S_{n-1}+\cdots +S_{n-p+1}) + v_n.
\end{eqnarray}
The noise terms $u_n$, $v_n$ and $w_n$ are assumed to be Gaussian white noise with
variances $\tau_1^2$, $\tau_2^2$ and $\sigma^2$, respectively (Kitagawa and Gersch (1984,1996) and 
Kitagawa (2020a)).

This seasonal adjustment model with two component models can be expressed in
state-space model form as
\begin{eqnarray}
x_n &=& F x_{n-1} + G v_n \nonumber \\
y_n &=& H x_n + w_n,
\end{eqnarray}
with $v_n \sim N(0,Q)$ and $w_n \sim N(0,1)$ and the state vector $x_n$ and 
the matrices $F$, $G$, $H$, $Q$ and $R$ are defined by
{\setlength{\arraycolsep}{1mm}
\begin{eqnarray}
x_n = \left[ \begin{array}{c} 
             T_n     \\
             T_{n-1} \\
             S_n     \\
             S_{n-1} \\
            \vdots\\ 
             S_{n-p+2} \end{array}\right],\quad
F &=& \left[ \begin{array}{cccccc} 
             2 &-1 &   &   &   &   \\
             1 & 1 &   &   &   &   \\
               &   &-1 &-1 &\cdots&-1\\
               &   & 1 &   &   &   \\
               &   &   &\ddots&&   \\ 
               &   &   &   & 1 &    \end{array}\right],\quad
G = \left[ \begin{array}{cc} 
             1 & 0   \\
             0 & 0  \\
             0 & 1  \\
             0 & 0  \\
          \vdots&\vdots\\ 
             0 & 0   \end{array}\right] \\
H &=& [\begin{array}{cccccc} 1&0&1&0&\cdots &0 \end{array}] \nonumber \\
Q &=& \left[ \begin{array}{cc} 
             \tau_1^2 & 0   \\
               0      & \tau_2^2 \end{array}\right], \quad
R = 1.
\end{eqnarray}  
}

It is noted that, similar to the trend model, it is possible to assume
that $R=1$.
In this case, the parameter is $(\tau_1^2,\tau_2^2)^T$,
and the $F$, $G$, $H$ and $R$ do not depend on the parameter.
In actual likelihood maximization, since there are positivity 
constrains, $\tau_1^2 > 0$ and $\tau_2^2 >0$,
we use the log-transformation,
\begin{eqnarray}
 \theta_1 = \log (\tau_1^2), \quad
 \theta_2 = \log (\tau_2^2) .
\end{eqnarray}

In this case, 
\begin{eqnarray}
 && \frac{\partial Q}{\partial\theta_1} 
    = \frac{\partial^2 Q}{\partial\theta_1\partial\theta_1} 
    = \left[ \begin{array}{cc} 
             \tau_1^2 & 0   \\
               0      & 0 \end{array}\right], \quad
  \frac{\partial Q}{\partial\theta_2} 
     = \frac{\partial^2 Q}{\partial\theta_2\partial\theta_2} 
     = \left[ \begin{array}{cc} 
               0 & 0   \\
               0 & \tau_2^2  \end{array}\right] \quad
  \frac{\partial^2 Q}{\partial\theta_1\partial\theta_2} 
     = \frac{\partial^2 Q}{\partial\theta_2\partial\theta_1} 
     = \left[ \begin{array}{cc} 
               0 & 0   \\
               0 & 0  \end{array}\right]  , \nonumber \\
 && \frac{\partial R}{\partial\theta_i} = 0, \quad
  \frac{\partial^2 R}{\partial\theta_i\partial\theta_j} = 0\quad (i,j=1,2). 
\end{eqnarray}
 
Since $F$, $G$ and $H$ do not depend on $\theta$ and $\displaystyle\frac{\partial F}{\partial\theta}=0$,
$\displaystyle\frac{\partial G}{\partial\theta}=0$ and $\displaystyle\frac{\partial H}{\partial\theta}=0$ hold,
we can use the same recursive algorithm for gradient and the Hessian of the log-likelihood 
as the one for the trend model.

For Whard  data, the standard seasonal adjustment model with $m_1=2$, $m_2=1$ is estimated
using the initial estimates of parameters, 
$\theta = (\log\tau_1^2, \log\tau_2^2)^T =(-5.29831, -14.98848)^T$.
The log-likelihood of the model with these initial parameters is $\ell (\theta )=-377.0849$
and the gradient and the Hessian obtained by the numerical difference and the proposed
method are shown in the Table \ref{Tab_TS_model}.
It can be seen that the numerical differentiation coincides with the analytic
derivative up to 5th digit.

The maximum likelihood estimates of the system noises are $\hat\tau^2_1 = 0.021185$,
$\hat\tau^2_2=0.0068434$ with $\ell (\hat\theta)=380.6569$.

\begin{table}[h]
\begin{center}
\caption{Comparison of the results by numerical diffference and the differential filter}\label{Tab_TS_model}

\vspace{2mm}
\begin{tabular}{c|cc}
        & Proposed method &  Numerical Difference \\
\hline
  Initial  &   &   \\
  $ \tau^2_0 $ & $ [ \begin{array}{cc}0.005000  & 0.0068160 \end{array} ]$ &
  $ [ \begin{array}{cc}0.005000  & 0.0068160 \end{array} ]$\\[1mm]
  $ \theta_0 $ & $ [ \begin{array}{cc}-5.29831  & -4.98848 \end{array} ]$ &
  $ [ \begin{array}{cc}-5.29831  & -4.98848 \end{array} ]$\\[2mm]
  $\ell (\theta )$ & $-377.0852$  &  $-377.0849$ \\[2mm]
  $\frac{\partial \ell(\theta )}{\partial \theta}$ & $\left[ 4.82627, -0.31490 \right]$ & $[4.82627, -0.31490]$ \\[2mm]
  $\frac{\partial^2 \ell(\theta )}{\partial \theta\partial\theta^T}$ & 
  $\left[ \begin{array}{cc} 3.17294 & -0.23066\\
                           -0.23066 &  0.41057 \end{array}\right] $  &
  $\left[ \begin{array}{cc} 3.17296 & -0.23066\\
                           -0.23066 &  0.41058 \end{array}\right] $ \\[4mm]
\hline
  Optimized   &  & \\
  $ \hat\tau^2$ & $ [\begin{array}{cc} 0.021185  &  0.0068434 \end{array} ]$ &
  $ [\begin{array}{cc} 0.021185  &  0.0068434 \end{array} ]$  \\[1mm]
  $ \hat\theta$ & $ [\begin{array}{cc} -3.8545  &  -4.9845 \end{array} ]$ &
  $ [\begin{array}{cc} -3.8545  &  -4.9845 \end{array} ]$  \\[2mm]
  $\ell (\theta )$ & 380.6569  &  380.6569 \\[2mm]
  $\frac{\partial \ell(\theta )}{\partial \theta}$ & $ [-2.3897\!\times \!\!10^{-5}\, -0.3747\!\times \!\!10^{-5}]$ & $[-1.46922\!\times \!\!10^{-6}\, -0.38865\!\times \!\!10^{-6}]$ \\[2mm]
  $\frac{\partial^2 \ell(\theta )}{\partial \theta \partial\theta^T}$ & 
  $\left[ \begin{array}{cc} 3.71968 & -0.20363\\
                           -0.20363 &  0.18135 \end{array}\right] $  &
  $\left[ \begin{array}{cc} 3.71968 & -0.20363\\
                           -0.20363 &  0.18135 \end{array}\right] $ \\[4mm]
\hline
\end{tabular}
\end{center}
\end{table}

Figure 3 shows the contour of the log-likelihood of the seasonal adjustment model.
The horizontal and the vertical axes indicate the common logarithms of $\tau^2_1$ and
$\tau^2_2$, respectively.
The log-likelihood has a similar values for smaller $\tau^2_2$ and form a platou.
The small value of $ \displaystyle\frac{\partial^2 \ell (\theta )}{\partial\theta_2\partial\theta_2} $,
0.18135, corresponds to this phenomenon.

\begin{figure}[tbp]
\begin{center}
\includegraphics[width=100mm,angle=0,clip=]{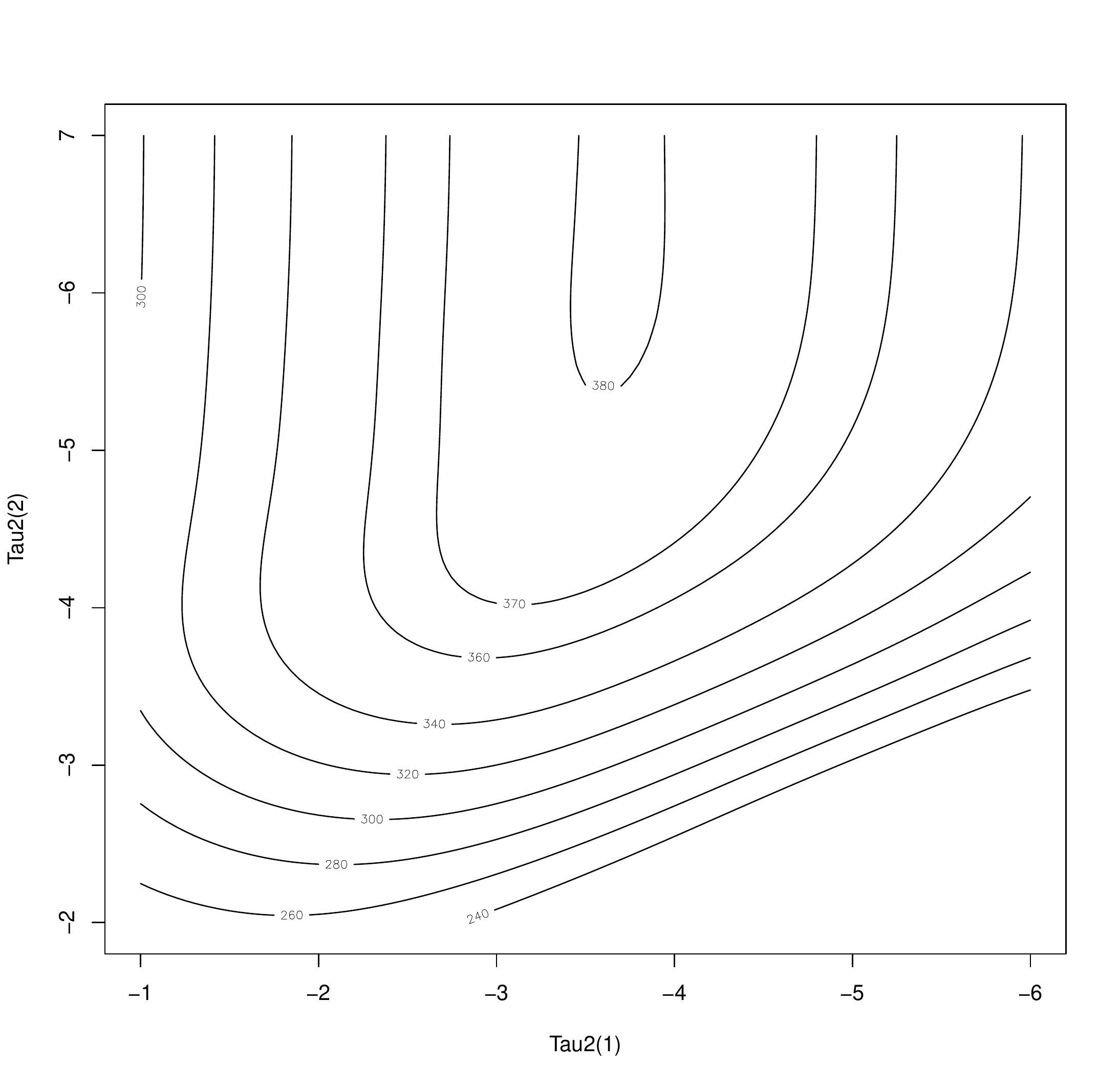}
\end{center}
\caption{Contour of log-likelihood function of seasonal adjustment model. Horizontal zxis: common logatithm of $\tau^2_1$,
Vertical axis: common logarithm of $\tau^2_2$. }
\label{Fig_contour of log-likelihood of TS_model.}
\vspace{5mm}
\end{figure}

\subsection{Seasonal adjustment model with stationary AR component}

The third example is a seasonal adjustment model with statinary AR component
\begin{eqnarray}
y_n = T_n + S_n + p_n +w_n,
\end{eqnarray}
where $T_n$ and $S_n$ are the trend component and the seasonal component
introduced in the previous subsection and $p_n$ is an AR component with AR order $m_3$ defined by
\begin{eqnarray}
 p_n = \sum_{j=1}^{m_3} a_j p_{n-j} + v_n^{(p)}. 
\end{eqnarray}
Here $v_n^{(p)}$is a Gaussian white noise with variance $\tau_3^2$.
The model contains $3+m_3$ parameters and the parameter vector is given by
$\theta =(\theta_1 ,\ldots ,\theta_{3+m_3})^T\equiv (\log\tau_1^2,\log\tau_2^2,\log\tau_3^2,
 \theta_4,\cdots ,\theta_{m_3+3})^T$.

The matrices $F$, $G$, $H$, $Q$ and $R$ are defined by
{\setlength{\arraycolsep}{1mm}
\begin{eqnarray}
x_n = \left[ \begin{array}{c} 
             T_n     \\
             T_{n-1} \\ \hline
             S_n     \\
             S_{n-1} \\
             \vdots  \\ 
             S_{n-p+2} \\ \hline
             p_{n-1}  \\
             p_{n-2}  \\
             \vdots   \\
             p_{n-m_3} \end{array}\right],\quad
F &=& \left[ \begin{array}{cc|cccc|cccc} 
             2 &-1 &   &   &   &   \\
             1 & 1 &   &   &   &   \\
             \hline
               &   &-1 &-1 &\cdots&-1\\
               &   & 1 &   &   &   \\
               &   &   &\ddots&&   \\ 
               &   &   &   & 1 &   \\
             \hline
               &   &   &   &   &   &a_1 & a_2 &\cdots & a_{m_3} \\
               &   &   &   &   &   & 1  &     &       &    \\
               &   &   &   &   &   &    & \ddots &    &    \\
               &   &   &   &   &   &    &        & 1  &
         \end{array}\right],\quad
G = \left[ \begin{array}{ccc} 
             1 & 0 & 0  \\
             0 & 0 & 0 \\  \hline
             0 & 1 & 0 \\
             0 & 0 & 0 \\
          \vdots&\vdots&\vdots\\ 
             0 & 0 & 0 \\  \hline
             0 & 0 & 1 \\
             0 & 0 & 0 \\
          \vdots&\vdots&\vdots\\ 
             0 & 0 & 0          
 \end{array}\right] \\
H &=& [\begin{array}{cccccccccc} 1&0&1&0&\cdots &0&1&0&\cdots &0 \end{array}] \nonumber \\
Q &=& \left[ \begin{array}{ccc} 
             \tau_1^2 & 0        & 0  \\
               0      & \tau_2^2 & 0  \\
               0      & 0  & \tau_3^2\end{array}\right], \quad
R = 1.
\end{eqnarray}  
}
The relation between the parameter $\theta_j$ and the variances and AR coefficients 
are as follows.
\begin{eqnarray}
&& \tau_j^2 = e^{\theta_j}, (j=1,\ldots ,3), \quad
  \beta_j = C\frac{e^{\theta_{j+3}}-1}{e^{\theta_{j+3}}+1}\\
&& \{a_j^{(m)} = a_j^{(m-1)} - \beta_m a_{m-j}^{(m-1)}, j=1,\ldots ,m \}, \quad \mbox{for }m=1,\ldots ,m_3.
\label{Eq_Levinson}
\end{eqnarray}
Note that the equation (\ref{Eq_Levinson}) is the relation between the AR coefficients 
of order $m-1$ and those  of the order $m$ used in the Levinson's algorithm (Kitagawa (2020a)).

In this case, 
\begin{eqnarray}
 && \frac{\partial Q}{\partial\theta_1} = \frac{\partial^2 Q}{\partial\theta_1\partial\theta_1}
    = \left[ \begin{array}{ccc} 
             \tau_1^2 & 0 & 0  \\
               0      & 0 & 0  \\
               0      & 0 & 0  \end{array}\right], \quad
  \frac{\partial Q}{\partial\theta_2} = \frac{\partial^2 Q}{\partial\theta_2\partial\theta_2}
    =\left[ \begin{array}{ccc} 
               0 & 0  & 0 \\
               0 & \tau_2^2 & 0 \\
               0 & 0  & 0\end{array}\right], 
 \nonumber\\
 && \frac{\partial Q}{\partial\theta_3} = \frac{\partial^2 Q}{\partial\theta_3\partial\theta_3}
    =\left[ \begin{array}{ccc} 
               0 & 0  & 0  \\
               0 & 0  & 0  \\
               0 & 0  & \tau_3^2\end{array}\right], \quad
    \frac{\partial^2 Q}{\partial\theta_i\partial\theta_j}
    =\left[ \begin{array}{ccc} 
               0 & 0  & 0  \\
               0 & 0  & 0  \\
               0 & 0  & 0  \end{array}\right], \quad \mbox{for }i \neq j \nonumber\\
 && \frac{\partial R}{\partial\theta_1} = 
    \frac{\partial R}{\partial\theta_2} = 
    \frac{\partial R}{\partial\theta_3} = 
    \frac{\partial^2 R}{\partial\theta_i\partial\theta_j} = 0,  (i,j=1,2,3)\\
&&\left(\frac{\partial F}{\partial \theta_k}\right)_{pq} =
   \left\{ \begin{array}{cl} \displaystyle
             \frac{\partial a_q^{(m)}}{\partial \theta_k} & \mbox{if }\left\{ \begin{array}{l}k=4,\ldots ,m_3+3, p=m_1+M_2(p-1)+1, \mbox{and}\\
                                                                q=m_1+M_2(p-1)+j, (j=1,\ldots ,m_3)\end{array}\right. \\
             0                                      & \mbox{otherwise} \end{array}
   \right. \nonumber \\
&&\left(\frac{\partial^2 F}{\partial \theta_j \partial \theta_k}\right)_{pq} =
   \left\{ \begin{array}{cl} \displaystyle
             \frac{\partial a_q^{(m)}}{\partial\theta_j \partial\theta_k} & \mbox{if }\left\{ \begin{array}{l}k=4,\ldots ,m_3+3, p=m_1+M_2(p-1)+1, \mbox{and}\\
                                                                q=m_1+M_2(p-1)+j, (j=1,\ldots ,m_3)\end{array}\right. \\
             0                                      & \mbox{otherwise} \end{array}
   \right. \nonumber
\end{eqnarray}
where $\displaystyle\left(\frac{\partial F}{\partial \theta_k}\right)_{pq}$ and 
$\displaystyle\left(\frac{\partial^2 F}{\partial\theta_j \partial\theta_k}\right)_{pq}$
denote the $(p,q)$ components
of the matrices $\displaystyle\frac{\partial F}{\partial \theta_k}$ and 
$\displaystyle\frac{\partial F^2}{\partial\theta_j \partial\theta_k}$,
respectively, and
$\displaystyle\frac{\partial a_i^{(m)}}{\partial\theta_j} $ and
$\displaystyle\frac{\partial^2 a_i^{(m)}}{\partial\theta_i\partial\theta_j} $ 
are obtained by
\begin{eqnarray}
\frac{\partial a_k^{(m)}}{\partial\theta_i} 
       &=&  \frac{\partial a_k^{(m)}}{\partial\beta_i} 
          \frac{\partial\beta_i}{\partial \theta_i} ,\quad i=1,\ldots ,m  \\
\frac{\partial^2 a_i^{(m)}}{\partial\theta_i \theta_j} 
       &=&  \frac{\partial^2 a_k^{(m)}}{\partial\beta_i \partial\beta_j} 
          \frac{\partial\beta_i}{\partial \theta_i} \frac{\partial\beta_j}{\partial \theta_j}
        + \frac{\partial a_k^{(m)}}{\partial\beta_i} 
          \frac{\partial^2 \beta_i}{\partial \theta_i \partial \theta_j} 
 ,\quad i,j=1,\ldots ,m\end{eqnarray}
and 
\begin{eqnarray}
 \frac{\partial\beta_k}{\partial \theta_i}
   &=& \left\{  \begin{array}{ll}
       \displaystyle C_{k}  & \mbox{for }k=i \\
       \displaystyle 0                                          & \mbox{for }k\neq i
       \end{array}  \right. \\
 \frac{\partial^2\beta_k}{\partial \theta_i\theta_j}
   &=& \left\{  \begin{array}{ll}
       \displaystyle D_{k}  & \mbox{for }k=i=j \\
       \displaystyle 0                                          & \mbox{otherwise}
       \end{array}  \right. \\
\frac{\partial a_k^{(m)}}{\partial\beta_i} 
    &=& \left\{  \begin{array}{ll}
          0         & \mbox{for }k=m \mbox{ and } i<m \\[3mm]
          1         & \mbox{for }k=m=i \\[2mm]
        \displaystyle \frac{\partial a_k^{(m-1)}}{\partial\beta_i}
            - \beta_m \frac{\partial a_{m-k}^{(m-1)}}{\partial\beta_i}
         & \mbox{for }k<m\mbox{ and }i <m \\[4mm]
           -  a_{m-k}^{(m-1)} 
         & \mbox{for }k<m\mbox{ and }i=m. \end{array}
        \right. \\
\frac{\partial^2 a_k^{(m)}}{\partial\beta_i \partial\beta_j} 
    &=& \left\{  \begin{array}{ll}
          0         & \mbox{for }k=m \\[3mm]
        \displaystyle \frac{\partial^2 a_k^{(m-1)}}{\partial\beta_i \partial\beta_j}
          - \frac{\partial\beta_m}{\partial \beta_j} \frac{\partial a_{m-k}^{(m-1)}}{\partial\beta_i}
          - \beta_m \frac{\partial^2 a_{m-k}^{(m-1)}}{\partial\beta_i \partial\beta_j}
         & \mbox{for }k<m\mbox{ and }i <m \\[4mm]
           -  \frac{\partial a_{m-k}^{(m-1)} }{\partial \beta_j}
         & \mbox{for }k<m\mbox{ and }i=m. \end{array}
        \right. 
\end{eqnarray}
In the above equations, $C_{i}$ and $D_{i}$ are the first and the second derivatives 
of the nonlinear transformation $\displaystyle C\frac{e^{\theta_i}-1}{e^{\theta_i}+1} $
and are given by
\begin{eqnarray}
  C_{i} = 2C \frac{e^{\theta_i}}{(e^{\theta_i}+1)^2}, \quad
  D_{i} = 2C \frac{e^{\theta_k}(1-e^{\theta_k})}{(e^{\theta_k}+1)^3} .
\end{eqnarray}

For $m_1=2$, $m_2=1$, the matrices $\displaystyle \frac{\partial F}{\partial \theta}$,
$\displaystyle \frac{\partial G}{\partial \theta}$, $\displaystyle \frac{\partial H}{\partial \theta}$, 
$\displaystyle \frac{\partial^2 F}{\partial \theta\partial \theta'}$, 
$\displaystyle \frac{\partial^2 G}{\partial\theta \partial\theta'}$ and 
$\displaystyle \frac{\partial^2 H}{\partial\theta \partial \theta'}$ are given by
\begin{eqnarray}
\frac{\partial F}{\partial \theta}
   = \left[ \begin{array}{ccc|c}
            0 & \ldots & 0 & 0 \\
            \vdots &\ddots & \vdots  & \vdots\\
            0 & \ldots & 0 & 0 \\
            \hline 
            0 & \ldots & 0 & \frac{\partial F_3}{\partial \theta}\rule{0mm}{5mm}
            \end{array}  \right], \quad
\frac{\partial G}{\partial \theta}
   = \left[ \begin{array}{ccc}
            0 & \ldots & 0  \\
            \vdots &\ddots & \vdots  \\
            0 & \ldots & 0  \\
            \end{array}  \right], \quad
\frac{\partial H}{\partial \theta}
   = \left[ \begin{array}{c} 0 \\ \vdots \\ 0 \end{array} \right]^T \\
\frac{\partial^2 F}{\partial\theta \partial\theta'}
   = \left[ \begin{array}{ccc|c}
            0 & \ldots & 0 & 0 \\
            \vdots &\ddots & \vdots  & \vdots\\
            0 & \ldots & 0 & 0 \\
            \hline 
            0 & \ldots & 0 & \frac{\partial^2 F_3}{\partial\theta \partial\theta'}\rule{0mm}{5mm}
            \end{array}  \right], \quad
\frac{\partial^2 G}{\partial\theta \partial\theta'}
   = \left[ \begin{array}{ccc}
            0 & \ldots & 0  \\
            \vdots &\ddots & \vdots  \\
            0 & \ldots & 0  \\
            \end{array}  \right], \quad
\frac{\partial^2 H}{\partial\theta \partial\theta'}
   = \left[ \begin{array}{c} 0 \\ \vdots \\ 0 \end{array} \right]^T
\end{eqnarray}
%
where for $m_3=1$, 2 and 3, $\displaystyle \frac{\partial F_3}{\partial \theta}$ and 
$\displaystyle \frac{\partial^2 F_3}{\partial\theta \partial\theta'}$ are respectively give by:
%

For $m_3=1$
\begin{eqnarray}
 F = \left[  a_1  \right], \qquad
\frac{\partial F}{\partial \theta_1} = 
  \left[ C_{1} \right], \quad
\frac{\partial^2 F}{\partial \theta_1\partial\theta_1} =
  \left[ D_{1} \right]. \nonumber 
\end{eqnarray}

For $m_3=2$
\begin{eqnarray}
 F &=& \left[ \begin{array}{cc} a_1 & a_2 \\[2mm] 1 & 0 \end{array} \right] \nonumber \\
\frac{\partial F}{\partial \theta_1} &=& 
  C_1 \left[ \begin{array}{cc}
        (1-\beta_2) & 0 \\[2mm]
         0 & 0
  \end{array} \right],\quad
\frac{\partial F}{\partial \theta_2} = 
  C_2 \left[ \begin{array}{cc}
        -\beta_1 & 1 \\[2mm]
        0 & 0
  \end{array} \right]
  \nonumber \\
\frac{\partial^2 F}{\partial \theta_1\partial\theta_1} &=& 
  D_1 \left[ \begin{array}{cc}
         (1- \beta_2) & 0 \\[2mm]
          0 & 0
  \end{array} \right], \quad
\frac{\partial^2 F}{\partial \theta_1\partial\theta_2} =
  -C_{1} C_{2} \left[ \begin{array}{cc}
        1 & 0 \\[2mm]
        0 & 0
  \end{array} \right] \nonumber \\
\frac{\partial^2 F}{\partial \theta_2\partial\theta_1} &=& 
  -C_{1} C_{2} \left[ \begin{array}{cc}
        1 & 0 \\[2mm]
        0 & 0
  \end{array} \right], \quad
\frac{\partial^2 F}{\partial \theta_2\partial\theta_2} =
  D_2 \left[ \begin{array}{cc}
       -\beta_1 &  1  \\[2mm]
        0 & 0
  \end{array} \right] \nonumber 
\end{eqnarray}

For $m_3=3$,
\begin{footnotesize}
\begin{eqnarray}
 F \!\! &=&\!\!  \left[ \begin{array}{ccc} a_1^{(3)} & a_2^{(3)} & a_3^{(3)} \\[2mm]
                                 1 & 0 & 0 \\[2mm]
                                 0 & 1 & 0 \end{array} \right] , \quad
\begin{array}{l}  a_1^{(3)} = a_1^{(2)} - a_3^{(3)}a_2^{(2)} = \beta_1-\beta_2\beta_1-\beta_3\beta_2, \\[4mm]
  a_2^{(3)} = a_2^{(2)} - a_3^{(3)}a_1^{(2)} = \beta_2 - \beta_3(\beta_1 - \beta_2\beta_1) \end{array}\nonumber \\
 \frac{\partial F}{\partial \theta_1}\!\!  &=& \!\! 
  C_1\! \left[ \tabcolsep=1mm\begin{tabular}{ccc}
    $\!1-\beta_2$ & $\beta_3(\beta_2 \!-\!1)$ & 0 \\[1mm]
     0 & 0 & 0 \\[1mm] 0 & 0 & 0
  \end{tabular} \right],
\frac{\partial F}{\partial \theta_2} = 
  C_2 \!\left[ \tabcolsep=1mm\begin{tabular}{ccc}
    $ \!-(\beta_1 +\beta_3)$ & $1+\beta_1\beta_3$ & 0 \\[1mm]
     0 & 0 & 0 \\[1mm] 0 & 0 & 0
  \end{tabular} \right], 
 \frac{\partial F}{\partial \theta_3} \!\! = 
  C_3 \!\left[ \tabcolsep=1mm\begin{tabular}{ccc}
    $ -\beta_2$ & $\beta_1(\beta_2-1)$ & 1 \\[1mm]
     0 & 0 & 0 \\[1mm] 0 & 0 & 0
  \end{tabular} \right]
  \nonumber \\
 \frac{\partial^2 F}{\partial \theta_1\partial\theta_1} \!\! &=& \!\! 
  D_{1}\!\! \left[ \begin{array}{ccc}
    \!\! 1\!-\!\beta_2 & \!\beta_3(\beta_2\!-\! 1) & 0\\[1mm] 0 & 0 & 0 \\[1mm] 0 & 0 & 0
  \end{array} \right], 
\frac{\partial^2 F}{\partial \theta_1\partial\theta_2} =
  C_1 C_2 \!\!\left[ \begin{array}{ccc} -1 & \beta_3 & 0 \\[1mm] 0 & 0 & 0 \\[1mm] 0 & 0 & 0 
  \end{array} \right] , 
\frac{\partial^2 F}{\partial \theta_1\partial \theta_3} = 
   C_1C_3 \!\left[ \begin{array}{ccc}
     0 & \!\beta_2\!-\!1 & 0 \\[1mm]
     0 & 0 & 0 \\[1mm] 0 & 0 & 0
  \end{array} \right]
\nonumber \\
 \frac{\partial^2 F}{\partial \theta_2\partial\theta_1} \!\! &=& \!\! 
  C_2 C_1 \!\!\left[ \begin{array}{ccc}
    -1 & \beta_3 & 0 \\[1mm]  0 & 0 & 0\\[1mm]  0 & 0 & 0
  \end{array} \right],
\frac{\partial^2 F}{\partial \theta_2\partial\theta_2} =
  D_{2}\!\!\left[ \begin{array}{ccc}\! \!-(\beta_1\!+\!\beta_3) & \!\!1+\!\! \beta_1\beta_3\! & 0 \\[1mm]  0 & 0 & 0\\[1mm]  0 & 0 & 0
  \end{array} \right] \nonumber , 
\frac{\partial^2 F}{\partial \theta_2 \partial \theta_3} = 
  C_2C_3 \!\!\left[ \begin{array}{ccc}
    -1 & \beta_1& 0 \\[1mm]
     0 & 0 & 0 \\[1mm] 0 & 0 & 0
  \end{array} \right]
\nonumber \\
 \frac{\partial^2 F}{\partial \theta_3\partial\theta_1} \!\! &=& \!\! 
 C_3 C_1 \!\!\left[ \begin{array}{ccc}
    0 & \!\beta_2\!-\!1\! & 0 \\[1mm]  0 & 0 & 0\\[1mm]  0 & 0 & 0
  \end{array} \right], 
\frac{\partial^2 F}{\partial \theta_3\partial\theta_2} =
 C_3 C_2 \!\!\left[ \begin{array}{ccc} -1 & \beta_1 & 0 \\[1mm]  0 & 0 & 0\\[1mm]  0 & 0 & 0
  \end{array} \right] \nonumber , 
\frac{\partial^2 F}{\partial \theta_3 \partial \theta_3} = 
   D_{3}\!\!\left[ \begin{array}{ccc}
     \!\!-\beta_2 & \!\!\beta_1(\beta_2\! - \! 1) & 1 \\[1mm]
     0 & 0 & 0 \\[1mm] 0 & 0 & 0
  \end{array} \right]
\nonumber
\end{eqnarray}
\end{footnotesize}


\vspace{3mm}
Since we have $
\displaystyle\frac{\partial G}{\partial\theta}=\displaystyle\frac{\partial H}{\partial\theta}
=\displaystyle\frac{\partial R}{\partial\theta}=0$
and $\displaystyle\frac{\partial^2 G}{\partial\theta\partial\theta^T}=
\displaystyle\frac{\partial^2 H}{\partial\theta\partial\theta^T}=
\displaystyle\frac{\partial^2 R}{\partial\theta\partial\theta^T}=0$
for the current model, the differential filter shown in 
(\ref{Eq_gradient_ell})-(\ref{Eq_Hessian_filter}) become considerably simple as follows.

\noindent
[The gradient of the log-likelihood]
\begin{eqnarray}
\frac{\partial\ell (\theta )}{\partial \theta} 
&=& - \frac{1}{2}\left( 
   \frac{N}{\hat\sigma^2}\frac{\partial \hat\sigma^2}{\partial\theta}
  + \sum_{n=1}^N  \frac{1}{r_n}\frac{\partial r_n}{\partial\theta}
   \right),  \label{Eq_gradient_ell_2}
\end{eqnarray}
where
\begin{eqnarray}
\frac{\partial \varepsilon_n}{\partial\theta} &=& -H \frac{\partial x_{n|n-1}}{\partial\theta} \nonumber \\
\frac{\partial r_n}{\partial\theta} &=& H \frac{\partial V_{n|n-1}}{\partial\theta}H^T, \\
\frac{\partial \hat\sigma^2}{\partial\theta} &=& \frac{1}{N}\sum_{n=1}^N  \left( \frac{2\varepsilon_n}{r_n} \frac{ \partial\varepsilon_{n}}{\partial\theta} 
 -  \frac{\varepsilon_{n}^2}{r_n^2} \frac{\partial r_n}{\partial\theta}  \right).
\end{eqnarray}

\noindent
[The derivative of the one-step-ahead predictor and the filter]
\begin{eqnarray}
\frac{\partial x_{n|n-1}}{\partial\theta} &=& F \frac{\partial x_{n-1|n-1}}{\partial\theta}
   + \frac{\partial F}{\partial\theta} x_{n-1|n-1}
\nonumber \\
\frac{\partial V_{n|n-1}}{\partial\theta} &=& F \frac{\partial V_{n-1|n-1}}{\partial\theta}F^T 
           + \frac{\partial F}{\partial\theta}V_{n-1|n-1}F^T 
           + F V_{n-1|n-1} \frac{\partial F}{\partial\theta}^T 
           + G\frac{\partial Q}{\partial\theta}G^T \\
\frac{\partial K_n}{\partial\theta} 
        &=& \frac{\partial V_{n|n-1}}{\partial\theta}H^T r_n^{-1} 
          - V_{n|n-1}H^T \frac{\partial r_n}{\partial\theta}r_n^{-2 }    \nonumber \\
\frac{\partial x_{n|n}}{\partial\theta} &=&  
        \frac{\partial x_{n|n-1}}{\partial\theta}
       + K_n \frac{\partial \varepsilon_n}{\partial\theta}
       + \frac{\partial K_n}{\partial\theta} \varepsilon_n   \nonumber \\
\frac{\partial V_{n|n}}{\partial\theta} &=& 
         \frac{\partial V_{n|n-1}}{\partial\theta}
       - \frac{\partial K_n}{\partial\theta} H V_{n|n-1}
       - K_n H \frac{\partial V_{n|n-1}}{\partial\theta}. \label{Eq_gradient-filter-F}
\end{eqnarray}
%
%
%


\noindent
[The Hessian of the log-likelihood] 
\begin{eqnarray}
\frac{\partial^2\ell (\theta )}{\partial \theta\partial \theta^T}
= - \frac{N}{2} \left( 
    \frac{1}{\hat\sigma^2}\frac{\partial^2 \hat\sigma^2}{\partial\theta\partial\theta^T} 
  - \frac{1}{(\hat\sigma^2)^2} \frac{\partial \hat\sigma^2}{\partial\theta}\frac{\partial \hat\sigma^2}{\partial\theta^T}  \right)   
-  \frac{1}{2}\sum_{n=1}^N  \left( 
 \frac{1}{r_n}\frac{\partial^2 r_n}{\partial\theta \partial\theta^T} 
- \frac{1}{r_n^2}\frac{\partial r_n}{\partial\theta }\frac{\partial r_n}{\partial\theta^T}  \right)
, \nonumber
\end{eqnarray}
where $ \displaystyle\frac{\partial^2 \varepsilon_n}{\partial\theta\partial\theta^T} $, 
$ \displaystyle\frac{\partial^2 r_n}{\partial\theta \partial\theta^T} $
and $ \displaystyle\frac{\partial^2 \hat\sigma^2_n}{\partial\theta \partial\theta^T} $  are
obtained by
\begin{eqnarray}
\frac{\partial^2 \varepsilon_n}{\partial\theta\partial\theta^T} 
&=&  -  H\frac{\partial^2 x_{n|n-1}}{\partial\theta\partial\theta^T}
 \nonumber \\
\frac{\partial^2 r_n}{\partial\theta \partial\theta^T} 
   &=&  H \frac{\partial^2 V_{n|n-1}}{\partial\theta \partial\theta^T} H^T \\
\frac{\partial^2 \hat\sigma^2}{\partial\theta \partial\theta^T} 
&=& \frac{1}{N}\sum_{n=1}^N  
\left\{ 
    \frac{2}{r_n}\!\!\left( \frac{ \partial \varepsilon_n}{\partial\theta} \frac{ \partial \varepsilon_n}{\partial\theta^T}
   +  \varepsilon_n \frac{\partial^2 \varepsilon_n}{\partial\theta \partial\theta^T} \right)
+  \frac{2\varepsilon_n^2}{r_n^3}  \frac{\partial r_n}{\partial\theta} \frac{\partial r_n}{\partial\theta^T} \right. \nonumber \\
 && {}\hspace{20mm}\left.
    -  \frac{\varepsilon_n}{r_n^2}\left( 2 \frac{\partial \varepsilon_n}{\partial\theta} \frac{\partial r_n}{\partial\theta^T}   
   +  2\frac{\partial r_n}{\partial\theta} \frac{\partial \varepsilon_n}{\partial\theta^T}  
   +  \varepsilon_n \frac{\partial^2 r_n}{\partial\theta \partial\theta^T} \right)    
 \right\} \nonumber 
\end{eqnarray}

\noindent
[The second derivatives of the one-step-ahead predictor and the filter]

\begin{eqnarray}
\frac{\partial^2 x_{n|n-1}}{\partial\theta \partial\theta^T}
   &=& \frac{\partial F}{\partial\theta^T} \frac{\partial x_{n-1|n-1}}{\partial\theta}
     + \frac{\partial F}{\partial\theta} \frac{\partial x_{n-1|n-1}}{\partial\theta^T}
     + F \frac{\partial^2 x_{n-1|n-1}}{\partial\theta \partial\theta^T}
     + \frac{\partial^2 F}{\partial\theta \partial\theta^T} x_{n-1|n-1}
\nonumber \\
\frac{\partial^2 V_{n|n-1}}{\partial\theta \partial\theta^T}
         &=& \frac{\partial F}{\partial\theta} \frac{\partial V_{n-1|n-1}}{\partial\theta^T}F^T  
          + \frac{\partial F}{\partial\theta^T} \frac{\partial V_{n-1|n-1}}{\partial\theta}F^T  
          + F \frac{\partial^2 V_{n-1|n-1}}{\partial\theta \partial\theta^T}F^T    \nonumber \\
      &&  + F \frac{\partial V_{n-1|n-1}}{\partial\theta}\frac{\partial F^T}{\partial\theta^T} 
          + F \frac{\partial V_{n-1|n-1}}{\partial\theta^T}\frac{\partial F^T}{\partial\theta} 
          + \frac{\partial^2 F}{\partial\theta \partial\theta^T}V_{n-1|n-1}F^T     \nonumber \\
      &&  +  \frac{\partial F}{\partial\theta}V_{n-1|n-1}\frac{\partial F^T}{\partial\theta^T} 
                         + \frac{\partial F}{\partial\theta^T}V_{n-1|n-1}\frac{\partial F^T}{\partial\theta} 
           + F V_{n-1|n-1} \frac{\partial^2 F^T}{\partial\theta \partial\theta^T} 
          + G\frac{\partial^2 Q}{\partial\theta \partial\theta^T}G^T  \nonumber \\
\frac{\partial^2 K_n}{\partial\theta \partial\theta^T} 
        &=& \frac{\partial^2 V_{n|n-1}}{\partial\theta \partial\theta^T}H^T  r_n^{-1}
        - \left(  \frac{\partial V_{n|n-1}}{\partial\theta}H^T \frac{\partial r_n}{\partial\theta^T} 
              +  \frac{\partial V_{n|n-1}}{\partial\theta^T}H^T \frac{\partial r_n}{\partial\theta} 
              + V_{n|n-1}H^T \frac{\partial^2 r_n}{\partial\theta \partial\theta^T}  \right) r_n^{-2}  \nonumber \\
       && + 2V_{n|n-1}H^T  \frac{\partial r_n}{\partial\theta} \frac{\partial r_n}{\partial\theta^T} r_n^{-3}\\
\frac{\partial^2 x_{n|n}}{\partial\theta \partial\theta^T} &=&  
        \frac{\partial^2 x_{n|n-1}}{\partial\theta \partial\theta^T}
       + \frac{\partial K_n}{\partial\theta} \frac{\partial \varepsilon_n}{\partial\theta^T}
       + \frac{\partial K_n}{\partial\theta^T} \frac{\partial \varepsilon_n}{\partial\theta} 
       + K_n \frac{\partial^2 \varepsilon_n}{\partial\theta \partial\theta^T} 
       + \frac{\partial^2 K_n}{\partial\theta \partial\theta^T} \varepsilon_n
\nonumber \\
\frac{\partial^2 V_{n|n}}{\partial\theta \partial\theta^T} &=& 
         \frac{\partial^2 V_{n|n-1}}{\partial\theta \partial\theta^T}
       - \frac{\partial^2 K_n}{\partial\theta \partial\theta^T} H V_{n|n-1}
       - \frac{\partial K_n}{\partial\theta} H \frac{\partial V_{n|n-1}}{\partial\theta^T} 
       - \frac{\partial K_n}{\partial\theta^T} H \frac{\partial V_{n|n-1}}{\partial\theta} 
     - K_n H \frac{\partial^2 V_{n|n-1}}{\partial\theta \partial\theta^T}. \nonumber
\end{eqnarray}

\begin{table}[h]
\begin{center}
\caption{Comparison of the gradient vectors and the Hessian matrix obtained by
the differential filter and the numerical differencing.}\label{Tab_TSAR_model_1}
\vspace{2mm}
By differential filter:\\
\begin{tabular}{lrrrrr}
 \hline
 $\frac{\partial\ell (\theta_0)}{\partial \theta}$ 
  &  3.472775 &  $-12.340431$ &     3.244577 &   21.235746 &    4.505459 \\
\hline
  &  0.320932 &    1.233847 &   $-2.574948$ &   $-4.783933$ &   $-0.262825$ \\
  &  1.233847 &  $-11.299148$ &    5.956742 &    7.917642 &    0.399997 \\
 $\frac{\partial^2\ell (\theta_0)}{\partial\theta \partial\theta^T}$ 
  & $-2.574948$ &    5.956742 &  $-10.176196$ &  $-11.233390$ &   $-0.372234$ \\
  & $-4.783933$ &    7.917642 &   11.233390 &  $-24.934335$ &    0.822854 \\
  & $-0.262825$ &    0.399997 &   $-0.372234$ &    0.822854 &    3.126537 \\
\hline
\end{tabular}
\vspace{2mm}

By numerical differencing:\\
\begin{tabular}{lrrrrr}
\hline
 $\frac{\partial\ell (\theta_0)}{\partial \theta}$ 
    &  3.472775 &  $-12.340431$ &    3.244577 &   21.235747  &   4.505459 \\
 \hline
    &  0.320913&     1.233881 &   $-2.574897$ &  $-4.783931$ &   $-0.262818$ \\
    &  1.233881 &  $-11.300011$ &    5.956570 &    7.917585 &    0.400087 \\
 $\frac{\partial^2\ell (\theta_0)}{\partial\theta \partial\theta^T}$ 
    & $-2.574897$ &    5.956570 &  $-10.176824$ &  $-11.233493$ &   $-0.372323$ \\
    & $-4.783931$ &    7.917585 &  $-11.233493$ &  $-24.934451$ &    0.822888 \\
    & $-0.262818$ &    0.400087 &   $-0.372323$ &    0.822888 &    3.126465 \\
\hline
\end{tabular}
\end{center}
\end{table}

Table \ref{Tab_TSAR_model_1} shows the gradients and the Hessian matrix obtained by 
the differential filter and the numerical differenting method. 
The initial estimates of the parameters are set to be
$\theta = (\log (0.25682\times 10^{-3}),\log( 1.0), \log( 0.52499),$ $1.7099, -0.89985)$
and the log-likelihood of the model is
$\ell (\theta )$ = $-348.9595$.

The maximum likelihoos estimates of the model are $\tau^2_1=1.8824\times 10^{-4}$,
$\tau^2_2=1.1348\times 10^{-2}$, $\tau^2_1=6.2550\times 10^{-2}$,
$a^{(2)}_1 = 1.6546$ and $a^{(2)}_2 = -0.6884$ with $\ell (\hat\theta)$ = 387.9554.
The gradients and the Hessian matrix for this maximum likelihood estimates are shown in
Table \ref{Tab_TSAR_model_2}.
In this case as well, the analytic derivative matches the numerical differentiation up to the fifth digit.

\begin{table}[tbp]
\begin{center}
\caption{Comparison of the gradient vectors and the Hessian matrix obtained by
the differential filter and the numerical differencing.}\label{Tab_TSAR_model_2}
\vspace{2mm}
By differential filter:\\
\begin{tabular}{lrrrrr}
 \hline
 $\frac{\partial\ell (\hat\theta)}{\partial \theta}$ 
   & 0.000001 &   $-0.000000$  &    0.000007 &    0.000000 &   $-0.000003$ \\
\hline
   &$-0.512604$ &    0.036523  &   $-0.186673$ &   $-0.000000$ &    0.029211 \\
   & 0.036523 &   $-0.305611$  &    0.371657 &    0.000000 &   $-0.349766$ \\
$\frac{\partial^2\ell (\hat\theta)}{\partial\theta \partial\theta^T}$ 
   &$-0.186673$ &    0.371657  &   $-8.066285$ &   $-0.000000$ &    7.186088 \\
   &$-0.000000$ &    0.000000  &   $-0.000000$ &   $-0.000000$ &   $-0.000000$ \\
   & 0.029211 &   $-0.349766$  &    7.186088 &   $-0.000000$ &  $-10.588254$ \\
\hline
\end{tabular}
\vspace{2mm}

By numerical differencing:\\
\begin{tabular}{lrrrrr}
\hline
 $\frac{\partial\ell (\hat\theta)}{\partial \theta}$ 
   & 0.000002 &   $-0.000001$ &    0.000011 &    0.000003 &   $-0.000002$ \\
\hline
   &$-0.512613$ &    0.036524 &   $-0.186672$ &   $-0.000001$ &    0.029217 \\
   & 0.036524 &   $-0.305652$ &    0.371642 &    0.000001 &   $-0.349770$ \\
$\frac{\partial^2\ell (\hat\theta)}{\partial\theta \partial\theta^T}$ 
   &$-0.186672$ &    0.371642 &   $-8.066400$ &   $-0.000003$ &    7.186086 \\
   &$-0.000001$ &    0.000001 &   $-0.000003$ &   $-0.000005$ &    0.000002 \\
   & 0.029217 &   $-0.349770$ &    7.186086 &    0.000002 &  $-10.588417$ \\
\hline
\end{tabular}
\end{center}
\end{table}


\section{Summary}
The gradient and the Hessian matrix of the log-likelihood of the reduced order linear state-space 
model are given. 
Details of the implementation of the algorithm for trend model, the standard seasonal adjustment model,
and the seasonal adjustment model with stationary AR component are given.
For each implementation, comparison with a numerical difference method is shown.

\vspace{15mm}
\noindent{\Large\bf Aknowledgements}

This work was supported in part by JSPS KAKENHI Grant Number 18H03210.
The author is grateful to the project members, Prof. Kunitomo, Prof Nakano,
Prof. Kyo, Prof. Sato, Prof. Tanokura and Prof. Nagao for their 
stimulating discussions.

\end{document}